\documentstyle[12pt]{article}
\begin{document}
\large
\centerline{\bf Vector Fields on a Disk with Mixed
Boundary Conditions}
\vspace{15mm}
\centerline{D.V.Vassilevich\footnote {e.-mail:
dvassil @ sph.spb.su}}
\vspace{5mm}

Department of Theoretical Physics, St.Petersburg
University,

198904 St.Petersburg, Russia

\vspace{10mm}

\centerline{Abstract}

\vspace{5mm}

We study vector fields on a disk satisfying two types
of boundary conditions. These boundary conditions are
selected by BRST-invariance in electrodynamics. They
also appear in the de Rham complex. The manifest
construction of the harmonic expansion is presented.
The eigenfunctions of the vector Laplace operator
are expressed in terms of fields satisfying pure
Dirichlet or Robin boundary conditions. For the case
of four-dimensional disk several first coefficients
of the heat kernel expansion are computed. An error
in the analitical expression by Branson and Gilkey is
corrected.

PACS: 02.40.-k, 03.50.De

\newline
SPbU-IP-94-6
\newline
gr-qc/9404052

\newpage

{\bf 1. Introduction}

\vspace{5mm}

Computations of the heat kernel expansion on a disk
goes back to the papers of Stewardson and Waecher [1].
One should also mention the subsequent results [2-5].
In the most complete form the analytic expressions for
the heat kernel expansion in terms geometric
characteristics of manifold were given by Branson and
Gilkey [6]. The modern interest to the problem is
related to calculations of pre-factor of the wave
function of the Universe [7-10].
Most of the above mentioned papers deal
with pure Dirichlet or Robin boundary conditions.
However, mixed boundary conditions appears naturally
in studying [6] the de Rham complex and in physical
applications [9,11]. Due to the facts that different
components of the fields obey different types of
boundary conditions and that the components satisfying
a certain type of boundary conditions do not belong to
an eigenspace of the Laplace operator, a direct study
of the heat kernel for mixed boundary conditions
become a very complicated task. Only very few examples
were studied.

In this paper we shall try to fill in the existing gap
by constructing the harmonic expansion for vector file
on a disk satisfying two types of mixed boundary
conditions and evaluating the heat kernel expansion.
These two types of boundary conditions are the same as
are selected by BRST-invariance [9] and that are
used in the de Rham complex [6]. To solve this problem
we first construct the Hodge-de Rham decomposition.
The transversal field is also decomposed in two
parts. This technique is similar to that used
previously [12] on homogeneous spaces. The proposed
decomposition is orthogonal and each part is an eigenspace
of the Laplace operator and can be described in terms
of fields satisfying pure Dirichlet or Robin boundary
conditions. In four dimensions the corresponding heat
kernel can be expressed entirely in terms of scalar
fields obeying pure boundary conditions. This
remarkable fact make it possible to compute several
first coefficients of the heat kernel expansion.

We also find an error in the analitic expression of
Branson and Gilkey [6] for the $a_2$ coefficient
for mixed boundary conditions.
This mistake is of technical nature. The authors [6]
overlooked a multiplier in one of the equations.
Since the paper [6] became one of the most
frequently cited sources on spectral geometry
of manifolds with boundary, we explain
how to correct this error in some detail. The
corrected expression is consistent with direct
computations on a disk.

The paper is organized is follows. In the next
section we build up harmonic expansion on
$d+1$-dimensional disk. Sec. 3 is devoted to the
heat kernel expansion. Computations for four
dimensional disk are presented. In Sec. 4 we examine
the analytic expressions for the Seeley coefficients.
Last section contains concluding remarks.

\vspace{5mm}

{\bf 2. The Hodge-de  Rham decomposition and harmonic
expansion}

\vspace{5mm}

Consider $d+1$ dimensional unit disk $M$ with the metric
$$ds^2=dr^2+d\Omega^2, \quad 0\le r \le 1 , \eqno (2.1)$$
where $d\Omega^2$ is the metric on unit sphere $S^4$.
The scalar Laplace operator has the form
$$\Delta \phi =
(\partial^2_0+\frac dr \partial_0+^{(d)}\Delta )\phi .
\eqno (2.2) $$
The $^(d)\Delta$ is the Laplace operator on $d$-dimensional
hypersurface. Throughout this paper we use notations $\{ x_\mu \}$
$=\{ x_0,x_i\}$, $x^0=r$, $\mu=$$0,1,...,d$. The eigenfunctions
of the operator (2.2) are well known
$$\phi_{l,\lambda }\propto r^{(1-d)/2 } J_{(d-1)/2 +l}
(r\lambda ) Y_{(l)}(x^i). \eqno (2.3)$$
The $J_p$ are the Bessel functions. The eigenvalues $-\lambda^2$
are defined by boundary conditions. $Y_{(l)}(x^i)$ are
$d$-dimensional scalar spherical harmonics. Their degeneracies
are
$$D_l^S=\frac {(2l+d-1)(l+d-2)!}{l!(d-1)!} . \eqno (2.4)$$

Consider now vector fields $A_\mu$. We shall construct the
Hodge-de Rham decomposition and give manifest description of
eigenfunctions of the Laplace operator. A vector field $A_\mu$
can be decomposed in transversal $A^T$ and longitudinal $A_L$
parts
$$A_\mu^L=\partial_\mu \phi , \quad \nabla^\mu A_\mu^T=0 ,
\eqno (2.5)$$
$\nabla^\mu$ is the covariant derivative. This decomposition
is orthogonal with respect to ordinary inner product
without surface term
$$<A,B>=\int_M d\mu (x) A_\nu (x)B^\nu (x) \eqno (2.6)$$
if the following surface integral vanishes
$$\int_{\partial M} d\mu (x) A_0 \phi =0 \eqno (2.7)$$
with $A_0=A_0^T$. This is true for one of the two types
of boundary conditions
$$A_0 \vert_{\partial M} =0 \quad {\rm or} \quad
\phi \vert_{\partial M} =0 \eqno (2.8)$$

The harmonic expansion for $A^L$ can be easily constructed
with the help of Eq. (2.5). Since the metric (2.1) is flat,
$$\Delta A_\mu^L (\lambda ,l)=\Delta \nabla_\mu
\phi_{\lambda ,l}=\nabla_\mu \Delta \phi_{\lambda ,l}
= -\lambda^2 A_\mu^L (\lambda , l). \eqno (2.9)$$
The fields $A^L$ corresponding to different $\lambda$ or
$l$ are orthogonal if the integral (2.7) vanishes.
This leads to one of the following boundary conditions
$$ \partial_0 \phi \vert_{\partial M} =0 \eqno (2.10a)$$
$$  \phi \vert_{\partial M} =0 \eqno (2.10b)$$
The first condition (2.10a) gives type $I$ boundary
condition for $A_\mu$
$$(I): \quad A_0 \vert_{\partial M} =0, \quad
\partial_0 A_i \vert_{\partial M} =
(\nabla_0 +1)A_i \vert_{\partial M} =0 \eqno (2.11)$$
For the second condition (2.10b) we have
$$(\partial_0+\frac dr )
A_0^L(\lambda ,l) \vert_{\partial M} =
(\partial_0 +\frac dr )\partial_0
\phi_{\lambda ,l}\vert_{\partial M} =
(\Delta -^{(d)}\Delta )\phi_{\lambda ,l}
\vert_{\partial M} = $$
$$=(-\lambda^2 +\frac {l(l+d-1)}{r^2} )
\phi_{\lambda ,l} \vert_{\partial M} = 0 \eqno (2.12)$$
The type $II$ boundary conditions are
$$(II): \quad (\partial_0 +d)A_0 \vert_{\partial M} = 0
, \quad A_i \vert_{\partial M} = 0 \eqno (2.13)$$
The eigenvalues are defined by equations (2.3) and (2.10).
Degeneracies are given by (2.4).

We naturally arrive at mixed boundary conditions (2.11) and
(2.13). In the context of de Rham complex they are known as
 absolute and relative boundary conditions [6]. In gauge
theories these  conditions are singled out by
BRST-invariance [9].

A bit more job is needed for manifest construction of
harmonic expansion of transversal fields. Let us recall
the vector Laplace operator with the metric (2.1).
$$(\Delta A)_0= (\partial_0^2 +\frac dr \partial_0
+^{(d)}\Delta - \frac d{r^2} )A_0 -\frac 2r ^{(d)}\nabla^i
A_i, $$
$$(\Delta A)_i= (\partial_0^2 + \frac {d-2}r \partial_0
+^{(d)}\Delta -\frac {d-1}{r^2} )A_i +\frac 2r \partial_iA_0.
\eqno (2.14)$$
The transversality condition reads
$$\nabla^\mu A_\mu =(\partial_0 +\frac dr )A_0
+^{(d)}\nabla^i A_i = 0 \eqno (2.15)$$
Consider first $d$-dimensional transversal vectors
$\ ^{(d)}\nabla^i A_i^\perp = 0$. The only non-singular
solution of (2.15) is $A_0=0$. After some algebra we
obtain the eigenfunctions of (2.14)
$$A_i^\perp \propto r^{(3-d)/2} J_{l+(d-1)/2}
(\lambda r)
Y_{(l)i}^\perp (x_i), \quad l=1,2,... \eqno (2.16)$$
with the eigenvalues $-\lambda^2$ defined by boundary conditions
(2.11) or (2.13). The $d$-dimensional transversal harmonics
$Y_{(l)}^\perp$ have the following degeneracies and
eigenvalues of $\ ^{(d)}\Delta$ [13]
$$D_l^\perp =
\frac {l(l+d-1)(2l+d-1)(l+d-3)!}{(d-2)! (l+1)!}, $$
$$\ ^{(d)}\Delta Y_{(l)i}^\perp =
-\frac 1{r^2} [l(l+d-1)-1] Y_{(l)i}^\perp \eqno (2.17)$$

For $d$-dimensional longitudinal fields solution of
$d+1$-dimensional transversality condition (2.15) can
be represented in the form
$$A_0(\psi )=^{(d)}\Delta r \psi , \quad
A_i (\psi )=-^{(d)}\nabla_i(\partial_0+\frac {d-2}r )
r \psi . \eqno (2.18)$$
$\psi$ is a scalar field obeying the decomposition (2.3)
where harmonics with $l=0$ are excluded. We used the operator
relation
$${\partial_0}^{(d)}\Delta =^{(d)}\Delta (\partial_0-
\frac 2r ).$$
One can check up by straightforward manipulations that
$$\Delta A_0(\psi )=^{(d)}\Delta r \Delta \psi ,$$
$$\Delta A_i (\psi )=-^{(d)}\nabla_i(\partial_0+\frac {d-2}r )
r \Delta \psi  \eqno (2.19)$$
and hence the spectrum of $\Delta$ on vector fields $A_\mu (\psi )$
is the same as of the scalar Laplacian on $\psi$. The boundary
conditions for the field $\psi$ are
$$(I): \quad \psi \vert_{\partial M}=0$$
$$(II): \quad (\partial_0 +\frac {d-1}r )
\psi \vert_{\partial M}=0. \eqno (2.20)$$
One can verify that the harmonics (2.18) are orthogonal for
orthogonal fields $\psi$ and for either type $I$ or type $II$
boundary conditions.

In this section we constructed an orthogonal decomposition for
vector field $A_\mu$ satisfying type $I$ or type $II$ mixed
boundary conditions
$$A_\mu=A_\mu^T +\partial_\mu \phi , \quad
A_\mu^T =A_\mu^\perp +A_\mu (\psi ) \eqno (2.21)$$
where all the fields $\phi$, $A_i^\perp$ and $\psi$
which define the spectrum of the vector Laplace operator
satisfy pure Dirichlet or Robin boundary conditions.

Note that the fields $A^\perp$ correspond to the so called
true physical variables of electrodynamics.

\vspace{5mm}

{\bf 3. Evaluation of the heat kernel}

\vspace{5mm}

The integrated heat kernel $K(t)$ can be represented as a sum
over spectrum
$$K(t)=\sum D_\lambda \exp (-t\lambda^2), \eqno (3.1)$$
where $-\lambda^2$ and $D_\lambda$ are eigenvalues of
$\Delta$ and their degeneracies. In the previous section
we defined the spectrum of the vector Laplace operator
with mixed boundary conditions in terms of fields
satisfying pure boundary conditions. According to
the decomposition (2.21) the $K(t)$ can be represented as
a sum of terms corresponding to different items in
(2.21). For the type $I$ boundary conditions (2.11),
(2.10a), (2.20) the integrated heat kernel is
$$K^I(t)=K_R(S=0;t)-1+K_D(t)-K_D^1(t)+K_R^\perp
(S=-1;t) \eqno (3.2)$$
where the superscript $R$ or $D$ corresponds to Robin
or Dirichlet boundary conditions. For the Robin
conditions
$$(\nabla_0-S)\Phi \vert_{\partial M}=0 \eqno (3.3)$$
we also give the value of $S$. The terms $K_R(t)-1$ in
(3.2) are the contribution of longitudinal fields. $K_R$
is the standard scalar heat kernel (see Appendix). The
term $-1$ is the subtraction of the mode $\phi=$const ,
which satisfies the boundary conditions but do not
generate longitudinal vector field. The $K_D-K_D^1$ is
the contribution of the fields $A(\psi )$. The term
$-K_D^1$  appeares due to the presence of the zero
modes $\psi^0$ : $\ ^{(d)}\nabla_i \psi^0=0$ in (2.18).
The last term corresponds to $A^\perp$.

For the type $II$ boundary conditions the integrated
heat kernel reads
$$K^{II}(t)=K_D(t)+K_R(S=1-d;t)-K_R^1(S=1-d;t)+K_D^\perp
(t) \eqno (3.4)$$
where the notations are obvious from the comments to the
equation (3.2). The mode $\phi=$const  is not allowed
by Dirichlet boundary conditions. For $d=3$ the last term
in (3.4) was evaluated by Louko [8].

As an example, consider the case $d=3$ which is interesting
due to applications to quantum cosmology. For the type $I$
boundary conditions we have using (2.16) and (2.17)
$$K_R^\perp (S=-1;t)=\sum_{n=2}^\infty \sum_{\lambda_n}
2(n^2-1) \exp (-\lambda^2_nt)=$$
$$=\sum_{n=1}^\infty \sum_{\lambda_n}
2(n^2-1) \exp (-\lambda^2_nt)= \eqno (3.5)$$
$$=2\sum_{n=1}^\infty \sum_{\lambda_n}
n^2\exp (-\lambda^2_nt)
-\sum_{n=0}^\infty \sum_{\lambda_n}
\exp (-\lambda^2_nt)+
\sum_{\lambda_0} \exp (-\lambda^2_0t)$$
where we have changed the summation index $n=l+1$. The eigenvalues
$\lambda_n$ are defined by the condition $J'_n(\lambda_n)=0$,
$J'_n(z)=\partial_z J_n(z)$. Consider the first sum in (3.5).
{}From (2.4) we see that $n^2$ is just $D^S_{n-1}$. The
condition $J_n(\lambda_n)=0$ is equivalent to
$$(\partial_0 +1) \frac 1r J_n (r\lambda_n)
\vert_{r=1} =0$$
Hence the first sum is just twice scalar heat kernel in
$d=3$ for $S=-1$. The second sum is also equivalent to
scalar heat kernel but in $d=1$.
$$K_R^{\perp }(S=-1;t)=2K_R(S=-1,d=3;t)-K_R(S=0,d=1;t)
+\sum_{\lambda_0} \exp (-\lambda^2_0t). \eqno (3.6)$$
Note that the last sum in (3.6) includes $\lambda_0=0$
because constant mode is also included in $K_R(S=0,d=1;t)$.

The term $K^1_D(t)$ is constructed from eigenfunctions of
the Laplace operator corresponding to the $l=0$ in (2.3).
$$K_D^1={\sum_\kappa}' \exp (-\kappa^2t), \eqno (3.7)$$
where the eigenvalues $\kappa^2$ are defined by the
condition $J_1(\kappa )=0$. Prime in (3.7) means that
zero eigenvalue is excluded by Dirichlet boundary
condition. Taking into account that ${J'}_0$$=-J_1$
we obtain
$$-{\sum_\kappa}' \exp (-\kappa^2t)-1+
\sum_{\lambda_0} \exp (-\lambda_0^2t)=0 \eqno (3.8)$$

Substituting (3.5)-(3.8) in (3.2) we arrive at the following
expression for the heat kernel
$$K^I(d=3;t)=K_R(S=0,d=3;t)+K_D(d=3;t)+2K_R(S=-1,d=3;t)$$
$$-K_R(S=0,d=1;t). \eqno (3.9)$$
All the terms in r.h.s. of (3.9) are scalar heat kernels
corresponding to pure boundary conditions. The small $t$
expansion can be obtained using standard expressions
(see Appendix).
$$K^I(d=3;t)=\frac 1{8t^2} +
\frac {\sqrt \pi}{8 t^{\frac 32}}
-\frac 1{4t}
-\frac {5 \sqrt \pi}{32 t^{\frac 12}} -\frac 1{45}
+O(t^{\frac 12}). \eqno (3.10)$$
For the type $II$ boundary conditions similar manipulations
with (3.4) give the following result
$$K^{II}(d=3;t)=3K_D(d=3;t)+K_R(S=-2,d=3;t)-K_D(d=1;t)
\eqno (3.11)$$
In the small $t$ limit we obtain
$$K^{II}(d=3;t)=\frac 1{8t^2} -
\frac {\sqrt \pi}{8 t^{\frac 32}}
-\frac 1{4t}
+\frac {25 \sqrt \pi}{128 t^{\frac 12}} -\frac {16}{45}
+O(t^{\frac 12}). \eqno (3.12)$$

Note that the sums like $K^1$ are cancelled in the final
expressions (3.9) and (3.11). However, they can be
easily evaluated by direct methods [1] based on the
Laplace transformation. In particular, our results
for small $t$ expansion of $K_D^1(d=3)$ agree with
previous ones [8].

\vspace{5mm}

{\bf 4. Analitic expressions}

\vspace{5mm}

In this section we review the method of Branson
and Gilkey [6], correct a minor error in their
calculations and compare the analitic expressions
with the result of previous section.

The aim of the paper [6] was to obtain analitic
expressions in terms of volume and boundary
integrals for the coefficints $a_k$ in the
small $t$ expansion for a generalized heat kernel
$${\rm Tr} (fe^{t\Delta })= t^{-\frac m2}
\sum_n t^n a_n(f,\Delta ), \quad m=d+1 \eqno (4.1)$$
where Tr is the functional trace, $f$ is arbitrary
function. Expression (4.1) coincides with (3.1) for
the special choice $f=1$.

A general expression for the coefficients $a_k$ corresponding
to mixed boundary conditions was derived [6](see Appendix).
This expression depends on several arbitrary parameters
$\beta$, which were defined using varios consistency
conditions. One of the conditions appeared due to
conformal variation of $\Delta$
$$\frac d{d\epsilon} \vert_{\epsilon =0}
a_n(1,e^{-2\epsilon f}\Delta )
-(m-2n)a_n(f, \Delta )=0 \eqno (4.2)$$
To make the Robin boundary conditions conformally
invariant the endomorphism $S$ was conformally
transformed in such a way to compensate the conformal
variation of the normal connection $\omega_N$, $N$ is
the inward pointing normal vector.The following variational
equations are usefull
$$\frac d{d\epsilon} \vert_{\epsilon =0} \nabla \Psi =0;
\quad \frac d{d\epsilon} \vert_{\epsilon =0}
k =-fk-df_{;N} ;$$
$$\frac d{d\epsilon} \vert_{\epsilon =0}  \omega_N=
\frac 12 (2-m)f_{;N};$$
$$\frac d{d\epsilon} \vert_{\epsilon =0}  S=
-fS+\frac 12 (m-2) f_{;N} \Pi_R \eqno (4.3)$$
For notations see Appendix.
In Ref. 6 the conformal variation of $S$ was written without
$\Pi_R$ in the right hand side. The necessity of $\Pi_R$ is
obvious, because variation of $S$ should compensate
variation of $\omega_N$ only on subspace $V_R$. Substituting
$a_2$ (see (A.3)) in (4.2) and collecting all terms with
$\Psi_{;i}\Psi^{;i}f_{;N}$ with the help of relation
 $tr (\Psi_{;i}\Psi^{;i} \Pi_R)$$=\frac 12 tr
 (\Psi_{;i}\Psi^{;i})$ we obtain
$$0=\frac 14 (m-2) \beta_5 -\beta_3 (m-1) -\beta_4
-\beta_6 (m-4) \eqno (4.4)$$
where the first term is $\frac 12$ times the corresponding
term in Ref. 6. Other relations were obtained [6] by
considerung the de Rham complex
$$\beta_5=-120, \quad 0=-48+2\beta_3+2\beta_4+120 \eqno (4.5)$$
Since the coefficients $\beta$ should be independent of $m$,
(4.4) and (4.5) can be easily solved giving the values of
$\beta$ listed in (A.5) instead of values (A.4) obtained in
Ref. 6. After substitution of geometric characteristics of
the disk in $d=3$ ($m=4$) we see, that the analitic expression
(A.3) is consistent with the expressions (3.10) and (3.12)
for our set of parameters (A.5). The old set (A.4) is inconsistent
with (3.10) and (3.12).  Note, that on two-dimensional disk
the $a_2$ depends on $\beta_3+\beta_4$ only , and the two sets
give equivalent results.

\vspace{5mm}

{\bf 5. Conclusions}

\vspace{5mm}

In this paper we constructed manfestly the harmonic expansion
for vector fields on a disk satisfying type $I$ or type $II$
mixed boundary conditions. These boundary conditions
ensure orthogonality of the Hodge-de  Rham decomposition
with respect to ordinary inner product. We calculated
several first coefficients of the heat kernel expansion for
the both types of boundary conditions on 4-dimensional disk.
A minor error in analitical expressins [6] was corrected.

The direct method suggested here can be generalized for other
spins. For example, one can consider gravitational field
on a disk with mixed boundary conditions which again are
favoured [9] by BRST - invariance. A much more tedious
task is related to bounded regions of curved space.
However, as far as several first Seeley coefficients
are concerned the corrected Branson and Gilkey expressions
(A.3) - (A.5) look quite relyable. It is highly inprobable
that any new errors will be found.

It is also interesting to extend the approach [14] to
to path integral over scalar fields in a cavity to the
case of vector fields and mixed boundary conditions.

\vspace{5mm}

{\bf Acknowledgment}

\vspace{3mm}

This work was supported by the Russian Foundation for
Fundamental Studies, grant \# 93-02-14378.

\newpage

{\bf Appendix}

\vspace{5mm}

This appendix contains the Seeley coefficients $a_k$
defined in Eq. (4.1) for Dirichlet, Robin and mixed
boundary conditions extarcted from the paper by
Branson and Gilkey [6]. For the Dirichlet boundary
conditions, $\Phi \vert_{\partial M}=0$, the $a_k$
are
$$a_0=(4\pi )^{-m/2}\int_M d\mu (x) {\rm tr}f,$$
$$a_{0.5}=-\frac 14 (4\pi )^{-d/2}
\int_{\partial M} d\mu (x) {\rm tr} f,$$
$$a_1=\frac 16 (4\pi )^{-m/2}\int_{\partial M} d\mu (x)
{\rm tr}
(2fk),$$
$$a_{1.5}=-\frac 14 (4\pi )^{-d/2} \frac 1{96}
\int_{\partial M} d\mu (x) {\rm tr}
[f(7k^2-10k_i^jk_j^i)],$$
$$a_2=\frac 1{360} (4\pi )^{-m/2}\int_{\partial M} d\mu (x)
{\rm tr} [f(24k_{;jj}+\frac {40}{21} k^3-\frac {88}7
k_i^jk_j^ik$$
$$+\frac {320}{21} k_i^jk_j^lk_l^i)], \quad m=d+1 \eqno (A.1)$$
We dropped all irrelevant terms proportional to derivatives
of $f$, Riemannn curvature, etc. $k_{ij}$ is the second
fundamental form, $k=k_i^i$, the ";" denotes  covariant
derivative. For the Robin boundary conditions,
$(\nabla_0-S)\Phi \vert_{\partial M}$$=0$, we have
$$a_0=(4\pi )^{-m/2}\int_M d\mu (x) {\rm tr}f,$$
$$a_{0.5}=\frac 14 (4\pi )^{-d/2}
\int_{\partial M} d\mu (x) {\rm tr} f,$$
$$a_1=\frac 16 (4\pi )^{-m/2}\int_{\partial M} d\mu (x)
{\rm tr}
[f(2k+12S)],$$
$$a_{1.5}=\frac 14 (4\pi )^{-d/2} \frac 1{96}
\int_{\partial M} d\mu (x) {\rm tr}
[f(13k^2+2k_i^jk_j^i+96Sk+192S^2)],$$
$$a_2=\frac 1{360} (4\pi )^{-m/2}\int_{\partial M} d\mu (x)
{\rm tr} [f(24k_{;jj}+\frac {40}3 k^3+8
k_i^jk_j^ik$$
$$+\frac {32}3 k_i^jk_j^lk_l^i+144Sk^2+48Sk_i^jk_j^i+480S^2k
+480S^3+120S_{;jj})], \eqno (A.2)$$.
Consider now mixed boundary conditions. Let $V=V_R\oplus V_D$
is a decomposition of tangent space parallel to normal
geodesics. $\Pi_R$ and $\Pi_D$ are projectors on $V_R$ and
$V_D$ respectively. Let $\Pi_R\Phi$ satisfies Robin
boundary conditions, and $\Pi_D \Phi$ satisfies Dirichlet
boundary conditions. Define $\Psi =\Pi_R-\Pi_D$.
$$a_0=(4\pi )^{-m/2}\int_M d\mu (x) {\rm tr}f,$$
$$a_{0.5}={\rm tr}\Psi a_{0.5}^R=
-{\rm tr}\Psi a_{0.5}^D,$$
$$a_1={\rm tr}(\Pi_Ra_1^R+\Pi_Da_1^D)$$
$$a_{1.5}={\rm tr}(\Pi_Ra_{1.5}^R+\Pi_Da_{1.5}^D)
+\frac 14 (4\pi )^{-d/2} \frac 1{96}
\int_{\partial M} d\mu (x) {\rm tr}
[f \beta_1 \Psi_{;j}\Psi^{;j}]$$
$$a_2={\rm tr}(\Pi_Ra_2^R+\Pi_Da_2^D)
+\frac 1{360} (4\pi )^{-m/2}\int_{\partial M} d\mu (x)
{\rm tr} [f(\beta_3 \Psi_{;j}\Psi^{;j}k+$$
$$+\beta_4 \Psi_{;j}\Psi_{;i}k^{ji}+
\beta_5 \Psi_{;j}\Psi^{;j} S)+...+
f_{;N}\Psi_{;j} \Psi^{;j}] \eqno (A.3)$$
where $a_k^D$ and $a_k^R$ are the Seeley coefficients
computed using Eqs. (A.1) and (A.2) respectively with
all traces resticted to $V_D$ and $V_R$. We dropped all
the terms with derivatives of $f$ exept for one which
is relevant for the analysis of section 4. $f_{;N}=
-\nabla_0 f$. Branson and Gilkey [6] give the following
values of coefficients $\beta$:
$$\beta_1=-12, \ \beta_3=-42, \
\beta_4=6, \ \beta_5=-120, \
\beta_6=-18. \eqno (A.4)$$
We argue that the correct values of $\beta_3$ and $\beta_4$
are
$$\beta_3=-12, \quad \beta_4=-24 \eqno (A.5)$$

\newpage

{\bf References}
\newline
1. K.Stewardson and R.T.Waecher, Proc. Cambridge Philos.
Soc. 69, 353
(1971);

R.T.Waecher, ibid. 72, 439 (1972).
\newline
2. G.Kennedy, J. Phys. A11, L173 (1978);

G.Kennedy, R.Critchley and J.S.Dowker, Ann. Phys. 125,
346 (1980).
\newline
3. I.G.Moss, Class. Quantum Grav. 6, 759 (1989).
\newline
4. I.G.Moss and J.S.Dowker, Phys. Lett. B229, 261 (1989);

J.Melmed, J.Phys. A21, L1131 (1989).
\newline
5. P.D.D'Eath and G.V.M.Esposito, Phys. Rev. D43, 3234
(1991); D44,
1713 (1991).
\newline
6. T.B.Branson and P.B.Gilkey, Commun. Part. Diff. Eqs
15, 245 (1990).
\newline
7. J.B.Hartle, Phys. Rev. D29, 2730 (1984);

K.Schleich, ibid. D32, 1889 (1985);

J.Louko, Ann. Phys. 181, 318 (1988).
\newline
8. J.Louko, Phys. Rev. D38, 478 (1988).
\newline
9. I.G.Moss and S.Poletti, Nucl. Phys. B341, 155 (1990).
\newline
10. I.G.Moss and S.Poletti, Phys. Lett. B245, 355 (1990).
\newline
11. H.C.Luckock and I.G.Moss, Class. Quantum Grav. 6, 1993
(1989).
\newline
12. D.V.Vassilevich, Lett. Math. Phys. 26, 147 (1992); Int.
J. Mod. Phys.
D2, 135 (1993).
\newline
13. M.A.Rubin and C.Ordonez, J.Math.Phys. 25, 2888 (1984).
\newline
14. L.Vanzo, J. Math. Phys. 34, 5625 (1993).

\end{document}